\documentclass[aps,showpacs,%amsmath,
amssymb,endfloats]{revtex4}
\usepackage{bm,amsfonts}
\begin{document}
\title {Light slowdown in the vicinity of cross-over resonances}
\author{M. Perdian, A. Raczy\'nski}\email{raczyn@phys.uni.torun.pl} \author{J. Zaremba}
\affiliation{Instytut Fizyki, Uniwersytet Miko\l aja Kopernika,
ulica Grudzi\c{a}dzka 5, 87-100 Toru\'n, Poland,}
%\email{raczyn@phys.uni.torun.pl}

\author{S. Zieli\'nska-Kaniasty}
\affiliation{Instytut Matematyki i Fizyki, Akademia
Techniczno-Rolnicza, Aleja Prof. S. Kaliskiego 7, 85-796
Bydgoszcz, Poland.}

\begin{abstract}
Pulse propagation is considered in an inhomogeneously broadened
medium of three-level atoms in a V-configuration, dressed by a
counter-propagating pump pulse. A significant signal slowdown is
demonstrated in this of the three frequency windows of a reduced
absorption and a steep normal dispersion, which is due to a
cross-over resonance. Particular properties of the group index in
the vicinity of such a resonance are demonstrated in the case of
closely spaced upper levels.

.
\\
\end{abstract}
\maketitle
\newpage
In recent years an interest has grown in studying light
propagation in atomic media having very special controlled optical
properties. The susceptibility of those media is steered by
control or pump laser fields to create the conditions of a lowered
absorption accompanied by a steep normal dispersion \cite{c1}. The
majority of works in this field have been done for three-level
atoms in the $\Lambda$ configuration, for which it was possible to
experimentally achieve \cite{c2,c3} and theoretically describe
\cite{c4,c5} light slowdown or even light stopping in the
conditions of the electromagnetically induced transparency
\cite{c6}. Reviews of various aspects of "slow light" concerning
both basic physical aspects as well as potential applications can
be found, e.g., in Refs \cite{d1,d2}.

Another situation, in which such a behavior of the susceptibility
is encountered, is typical of the Doppler-free saturated
spectroscopy \cite{c7}. As recently pointed out by Agarwal and Dey
\cite{c8}, in a region of the Lamb dip, created by a strong laser
beam in a two-level system, conditions of a reduced absorption and
steep dispersion are created for a counter-propagating signal
pulse, being in resonance with the only transition in this system.

In this work we make a step further and study light propagation in
an inhomogeneously broadened medium of three-level atoms in the V
configuration. Although general optical properties of such media
are well known, as concerns the possible enhancement of the index
of refraction there exists a completely new element compared with
the case of two-level systems. In addition to two resonances due
to a resonant coupling between two states, there appears a
cross-over resonance \cite{c7} and a region of a peculiar behavior
of the susceptibility connected with it. In this region one
observes a significant reduction of the absorption and a steep
normal dispersion. We thus have to do with a kind of a
transparency window, in which the group velocity is reduced but,
in contrast to the case of the $\Lambda$ configuration, the
absorption is not completely suppressed. Note that cross-over
resonances have recently been discussed in the context of double
dark resonances appearing in generalized $\Lambda$ systems with an
additional coupling to a fourth level \cite{c9}.

In this paper we will investigate the pulse propagation in the
vicinity of a cross-over resonance. In particular we will evaluate
the scale of the effect of the pulse slowdown and its dependence
on the energy spacing of the upper levels.

Consider a three-level model atom with a single lower state $b$
and two upper states $a$ and $d$. The lower state is coupled with
the upper states by two counter-propagating fields: the weak
signal field $\frac{1}{2}\epsilon_{1}(z,t)
\exp[i(k_{1}z-\omega_{1}t)]+\frac{1}{2} \epsilon_{1}^{*}(z,t)
\exp[-i(k_{1}z-\omega_{1}t)]$ (the phase of the signal changes
during the evolution) and a relatively strong control field
$\epsilon_{2}\cos(k_{2}z-\omega_{2}t)$ ($k_{2}<0$). The density
matrix $\rho$ fulfills the von Neumann equation completed with the
phenomenological relaxation terms describing relaxation within the
system. If we transform-off the terms rapidly oscillating in time
and make the rotating wave approximation we obtain the following
equations for the density matrix $\sigma$
($\sigma_{ba}=\rho_{ba}\exp[-i(k_{2}z-\omega_{2}t)]$,
$\sigma_{bd}=\rho_{bd}\exp[-i(k_{2}z-\omega_{2}t)]$,
$\sigma_{ij}=\rho_{ij}$ otherwise)
\begin{eqnarray}
i\hbar\dot{\sigma}_{aa}=&&
-d_{ab}\sigma_{ba}\{\frac{1}{2}\epsilon_{1}
\exp[(i(k_{1}-k_{2})z-i(\omega_{1}-\omega_{2})t]+\frac{1}{2}\epsilon_{2}\}+
\nonumber\\&&d_{ba}\sigma_{ab}\{\frac{1}{2}\epsilon_{1}^{*}
\exp[i(k_{2}-k_{1})z-i(\omega_{2}-\omega_{1})t]+\frac{1}{2}\epsilon_{2}\}
-i\hbar\Gamma_{aa}\sigma_{aa},\nonumber\\
i\hbar\dot{\sigma}_{dd}=&&
-d_{db}\sigma_{bd}\{\frac{1}{2}\epsilon_{1}
\exp[(i(k_{1}-k_{2})z-i(\omega_{1}-\omega_{2})t]+\frac{1}{2}\epsilon_{2}\}+
\nonumber\\&&d_{bd}\sigma_{db}\{\frac{1}{2}\epsilon_{1}^{*}
\exp[i(k_{2}-k_{1})z-i(\omega_{2}-\omega_{1})t]+\frac{1}{2}\epsilon_{2}\}
-i\hbar\Gamma_{dd}\sigma_{dd},\nonumber\\
i\hbar\dot{\sigma}_{ba}=&&(E_{b}+\hbar\omega_{2}-E_{a})\sigma_{ba}+
[d_{ba}(\sigma_{bb}-\sigma_{aa})-d_{bd}\sigma_{da}]
\{\frac{1}{2}\epsilon_{1}^{*}\exp[i(k_{2}-k_{1})z-
i(\omega_{2}-\omega_{1})t]+\frac{1}{2}\epsilon_{2}\}-i\hbar\Gamma_{ba}\sigma_{ba},\\
i\hbar\dot{\sigma}_{bd}=&&(E_{b}+\hbar\omega_{2}-E_{d})\sigma_{bd}+
[d_{bd}(\sigma_{bb}-\sigma_{dd})-d_{ba}\sigma_{ad}]
\{\frac{1}{2}\epsilon_{1}^{*}\exp[i(k_{2}-k_{1})z-
i(\omega_{2}-\omega_{1})t]+\frac{1}{2}\epsilon_{2}\}
-i\hbar\Gamma_{bd}\sigma_{bd},\nonumber\\
i\hbar\dot{\sigma}_{ad}=&&(E_{a}-E_{d})\sigma_{ad}
-d_{ab}\sigma_{bd}\{\frac{1}{2}\epsilon_{1}\exp[i(k_{1}-k_{2})z
-i(\omega_{1}-\omega_{2})t]+\frac{1}{2}\epsilon_{2}\}
+\nonumber\\&&d_{bd}\sigma_{ab}\{\frac{1}{2}\epsilon_{1}^{*}
\exp[i(k_{2}-k_{1})z-(\omega_{2}-\omega_{1})t]+
\frac{1}{2}\epsilon_{2}\}-i\hbar\Gamma_{ad}\sigma_{ad},\nonumber
\end{eqnarray}
where $d_{ij}$ are the dipole moment matrix elements and
$\Gamma$'s - the relaxation rates. Equations (1) are solved
perturbatively with respect to the signal field $\epsilon_{1}$,
i.e. we write $\sigma_{ij}=\sigma_{ij}^{0}+\sigma_{ij}'$, where
$\sigma_{ij}^{0}$ are the solutions (stationary) of Eqs (1)
without the signal field, while $\sigma_{1}'$ are linear in
$\epsilon_{1}$. In the Fourier picture the equations for
$\sigma_{ij}'(\omega)$ read

\begin{eqnarray}
(\omega+i\Gamma_{aa})\sigma_{aa}'+\frac{1}{2\hbar}d_{ab}\epsilon_{2}\sigma_{ba}'-
\frac{1}{2\hbar}d_{ba}\sigma_{ab}'=&&-\frac{1}{2\hbar}d_{ab}\sigma_{ba}^{0}
\exp[i(k_{1}-k_{2})z]\epsilon_{1}(\omega-\omega_{1}+
\omega_{2})+\nonumber\\&& \frac{1}{2\hbar}d_{ba}\sigma_{ab}^{0}
\exp[i(k_{2}-k_{1})z]\epsilon_{1}^{*}(-\omega+\omega_{1}-\omega_{2}),\nonumber\\
(\omega+i\Gamma_{dd})\sigma_{dd}'+\frac{1}{2\hbar}d_{db}\epsilon_{2}\sigma_{bd}'-
\frac{1}{2\hbar}d_{bd}\sigma_{db}'=&&-\frac{1}{2\hbar}d_{db}\sigma_{bd}^{0}
\exp[i(k_{1}-k_{2})z]\epsilon_{1}(\omega-\omega_{1}+\omega_{2})+\nonumber\\&&
\frac{1}{2\hbar}d_{bd}\sigma_{db}^{0}
\exp[i(k_{2}-k_{1})z]\epsilon_{1}^{*}(-\omega+\omega_{1}-\omega_{2}),\nonumber\\
(\omega+\omega_{ab}-\omega_{2}+i\Gamma_{ba})\sigma_{ba}'+\frac{1}{2\hbar}d_{ba}
\epsilon_{2}(\sigma_{aa}'-\sigma_{bb}')&&+\frac{1}{2\hbar}d_{bd}\epsilon_{2}
\sigma_{da}'=\\ {[
d_{ba}(\sigma_{bb}^{0}-\sigma_{aa}^{0})-d_{bd}\sigma_{da}^{0} ]}&&
\frac{1}{2\hbar}
\exp[i(k_{2}-k_{1})z]\epsilon_{1}^{*}(-\omega+\omega_{1}-\omega_{2}),\nonumber\\
(\omega+\omega_{db}-\omega_{2}+i\Gamma_{bd})\sigma_{bd}'+\frac{1}{2\hbar}d_{bd}
\epsilon_{2}(\sigma_{dd}'-\sigma_{bb}')&&+\frac{1}{2\hbar}d_{ba}\epsilon_{2}
\sigma_{ad}'=\nonumber\\
{[d_{bd}(\sigma_{bb}^{0}-\sigma_{dd}^{0})-d_{ba}\sigma_{ad}^{0}]}&&\frac{1}{2}
\exp[i(k_{2}-k_{1})z]\epsilon_{1}^{*}(-\omega+\omega_{1}-\omega_{2}),\nonumber\\
(\omega-\omega_{ad}+i\Gamma_{ad})\sigma_{ad}'+
\frac{1}{2\hbar}d_{ab}\epsilon_{2}\sigma_{bd}'-
\frac{1}{2\hbar}\epsilon_{2}d_{bd}\sigma_{ab}'=&&\nonumber\\
-\frac{1}{2\hbar}d_{ab}\sigma_{bd}^{0}\exp[i(k_{1}-k_{2})z]
\epsilon_{1}(\omega-\omega_{1}+\omega_{2})+&&\frac{1}{2\hbar}d_{bd}\sigma_{ab}^{0}
\exp[i(k_{2}-k_{1})z]\epsilon_{1}^{*}(-\omega+\omega_{1}-\omega_{2}),\nonumber
\end{eqnarray}
where $\omega_{ij}=(E_{i}-E_{j})/\hbar$. The equations for the
other nondiagonal matrix elements follow from the relation
$\sigma_{ij}(\omega)=\sigma_{ji}^{*}(-\omega)$. Note that
$\sigma_{aa}^{0}+\sigma_{bb}^{0}+\sigma_{cc}^{0}=1$,
$\sigma_{aa}'+\sigma_{bb}'+\sigma_{cc}'=0$.

The polarization of the medium is $P(t)=P^{+}(t)+P^{-}(t)$ where
\begin{equation}
P^{+}(t)=N [d_{ba}\sigma'_{ab}(t)+d_{bd}\sigma'_{bd}(t)]
\exp[i(k_{2}-k_{1})z-i(\omega_{2}-\omega_{1})t],
\end{equation}
$N$ being the medium density. Alternatively we may write
\begin{equation}
P^{+}(\omega)=N [d_{ba}\sigma'_{ab}(\omega+\omega_{1}-\omega_{2})+
d_{bd}\sigma'_{db}(\omega+\omega_{1}-\omega_{2})]\exp[i(k_{2}-k_{1})z].
\end{equation}
This means that in the case of counter-propagating fields the
contribution of the terms including $\epsilon^{*}_{1}$ in Eqs (2)
is averaged to zero due to rapid spatial oscillations. The
susceptibility is given by the formula
$\chi(\omega,\omega_{1},\omega_{2})=
\frac{2}{\epsilon_{0}\epsilon_{1}(\omega)}
P(\omega)\exp[i(k_{1}-k_{2})z]$, $\epsilon_{0}$ being the vacuum
permittivity (the factor of 2 is due to the fact that we have
included the factor of 1/2 in the definition of the electric
fields). Because of the inhomogeneous broadening of the medium the
calculated susceptibility is to be averaged over the atomic
velocity distribution of the width $D$
\begin{equation}
\chi_{av}(\omega)=\int_{-\infty}^{\infty}
\chi(\omega,\omega_{1}-k_{1}v,\omega_{2}-k_{2}v)\frac{1}{\sqrt{\pi}D}
\exp[-\frac{v^{2}}{D^{2}}] dv.
\end{equation}

Due to the susceptibility being a rapidly varying function of
$\omega$ the group velocity $v_{g}$ differs from $c$ by the group
index $n_{g}=c/v_{g}$ of a large value
\begin{equation}
n_{g}=1+\frac{\omega_{1}}{2}\frac{d}{d\omega} {\rm Re}
\chi_{av}(\omega).
\end{equation}

The propagation inside the sample of a light pulse of a spectral
shape $g(\omega)$ is given by
\begin{equation}
\epsilon_{1}(z,t)=\int_{-\infty}^{\infty} g(\omega)
\exp[-i(\omega(t-\frac{z}{c})]\exp[i\frac{\omega_{1}z}{2c}\chi_{av}(\omega)]d\omega.
\end{equation}

For the sake of illustration we have performed numerical
calculations of the susceptibility by finding a numerical solution
of the stationary version of Eqs (1) with $\epsilon_{1}$=0 and
then by solving Eqs (2). Because our discussion of the influence
of cross-over resonances on laser pulse propagation has a rather
general character we have adopted somewhat arbitrarily the input
data being of order of those corresponding to rubidium and its
hyperfine splitting. The obtained susceptibility was averaged with
the Gausssian distribution with $k_{1}D=4.58\times10^{-8}$ a.u.
which corresponds to a Doppler FWHM of $\Delta\nu$=502 MHz. The
density of the medium was $3\times10^{-14}$ a.u. ($2\times10^{11}$
cm$^{-3}$). The atomic parameters were taken
$\omega_{da}=4.06\times10^{-8}$ a.u. ($\nu$=267 MHz),
$\omega_{ab}=5.845\times10^{-2}$ a.u. ($\lambda=780$ nm),
$\Gamma_{aa}=\Gamma_{dd}\equiv \Gamma=8.63\times 10^{-10}$ a.u.
($\delta\nu=$5.7 MHz),
$\Gamma_{ba}=\Gamma_{bd}=\Gamma_{ad}/2=\Gamma/2$, which
corresponds to a spontaneous emission from both upper levels of a
lifetime of 28 ns (a generalization for the case of different
lifetimes of the two upper levels is straightforward). The matrix
elements of the dipole moments were obtained from the values of
the relaxation rates, assuming that the spontaneous emission
occurs only to the lower state $|b>$. The maximum pump field
amplitude was $3.47\times10^{-10}$ a.u. (4.23 mWcm$^{-2})$. We
have set $\omega_{1}=\omega_{2}$ ($k_{2}=-k_{1}$),
$\hbar\omega_{1}=(E_{a}+E_{d})/2-E_{b}$. Our scanning of the
Fourier variable is equivalent to an independent scanning with the
probe laser, as usual in the studies of electromagnetically
induced transparency.

Fig. \ref{f1} presents the group index as a function of frequency.
The left and right maxima correspond to those frequencies for
which one observes a steep normal dispersion and a reduced
absorption due to the excitation by the probe pulse of the lower
and upper excited levels, respectively, in two groups of atoms of
such velocities that their number in the ground state has been
decreased because of transitions to the same level due to the
pump. Each of the two maxima is an analogue of the single maximum
observed in Ref. \cite{c8}. Our central maximum is a new
phenomenon: we observe a reduction of the group velocity,
connected with a steep dispersion and a reduced absorption in the
frequency region which corresponds to a cross-over resonance, i.e.
to an excitation of atoms from the holes of the velocity
distribution to the upper states different from those which took
part in the hole burning. The group index due to the cross-over
resonance, similarly as that due to two-level resonances,  can be
as large as a few hundreds even for a small atomic density.

If the lasers are not tuned  exactly in the middle between the
upper levels the three peaks are shifted and made asymmetric,
because the population of the atoms of such a velocity that they
can be active in the transitions is changed.

Fig. \ref{f2} presents the shape of an initially Gaussian pulse
inside the sample, propagating in the window of a reduced
absorption around the cross-over resonance, obtained from Eq. (7).
At the end of the 0.5 cm sample one observes a reduction of the
height to the level of 3 \% of the initial value due to the pulse
absorption, which is a typical value in saturation absorption
experiments, and a retardation by $4\times10^{8}$ a.u. (about 10
ns) due to the decrease of its group velocity.

Besides V systems composed of atomic hyperfine levels it is
possible to realize such a system when the upper levels are chosen
to be Zeeman sublevels of an excited level with properly chosen
polarizations of the laser beams. The latter case allows for an
additional control of the dispersive properties of the medium by
changing the level spacings using an external magnetic field.

It is interesting to  study the situation in which the energy
spacing of the close upper levels is of order of the natural
width. Fig. \ref{f3} shows the frequency dependence of the group
index for different level spacings. Curve 1, corresponding to
$\omega_{da}=4\Gamma$ exhibits again three maxima, however with
the central one lowered due a partial overlapping of the two holes
burned out in the velocity distribution. Reducing the level
spacing to $2\Gamma$ (curve 2) caused disappearing of the central
peak, because the gap between the two holes has vanished. Merging
the levels leads to a single peak (curve 3). This corresponds to a
single dip in the absorption spectrum. It follows from the Bloch
equations that such a situation is equivalent to the case of a
two-level system, in which the upper state is a certain "bright"
combination of $|a>$ and $|d>$ (the other orthogonal combination
being decoupled from $|b>$), the transition dipole moment is
$\sqrt{|d_{ba}|^{2}+|d_{bd}|^{2}}$ and the dip is the Lamb dip.
The widths of the holes depend on the intensity of the control
field. Curve 1 in Fig.\ref{f4} is the same as curve 2 in Fig.
\ref{f3}. Decreasing the field intensity leads to reappearing of
the central peak (curve 2), because the holes have been narrowed
and thus separated again. Increasing $\epsilon_{2}$ causes such a
broadening of the holes that they overlap, which yields the group
index in the form of a single broad peak. The dependence of the
heights of the peaks on the field intensity is connected with the
fact that in weaker fields the holes are narrower, so are the
transparency windows and the regions of a normal dispersion; as a
consequence the dispersion curves are steeper and the group index
is larger (Fig. \ref{f3}).

We have demonstrated the possibility of a significant enhancement
of the group index and a slowdown of a light pulse in an
inhomogeneously broadened medium of three-level atoms in the
V-configuration. The Doppler-averaged susceptibility exhibits
three regions of a reduced absorption accompanied by a steep
normal dispersion. We have examined laser pulse propagation in the
central region corresponding to a cross-over resonance. The pulse
slowdown is of the same order of magnitude as in the two remaining
regions. In the case of closely spaced upper levels of a V system
we have demonstrated how the frequency dependence of the group
index changes depending on the level spacing and the pump field
intensity.

\begin{acknowledgments}
This work is a part of a program of the National Laboratory of AMO
Physics in Toru\'n, Poland.
\end{acknowledgments}

\newpage

\begin{figure}
\caption{\label{f1} The group index for the data given in the
text. $10^{-8}$ a.u. on the frequency axis corresponds to 12
$\Gamma$.}
%\label{f2}
\end{figure}

\begin{figure}
\caption{\label{f2} The shape of a signal pulse along the sample;
left: from top to bottom at $z=0$, $z=2\times 10^{7}$ a.u.,
$z=6\times10^{7}$ a.u., $z=10^{8}$ a.u. (0.53 cm) for the data as
in the text; right: normalized pulse peak at the end of the sample
(dashed line) compared with an analogous pulse traveling in vacuum
(solid line). $10^{10}$ a.u. on the time axis is equal to
2.4$\times10^{-7}$ s and corresponds to 8.63 $\Gamma^{-1}$.}
%\label{f3}
\end{figure}

\begin{figure}
\caption {\label{f3} The group index in the case of closely spaced
upper levels for $\epsilon_{2}=3.47\times10^{-10}$ a.u.:
$\omega_{da}=4\Gamma$ (line 1), $\omega_{da}=2\Gamma$ (line 2),
$\omega_{da}=0$ (line 3). $10^{-9}$ a.u. on the frequency axis
corresponds to 1.2 $\Gamma$. }
%\label{f3}
\end{figure}

\begin{figure}
\caption {\label{f4} The group index in the case of closely spaced
upper levels for $\omega_{da}=2\Gamma$:
$\epsilon_{2}=1.74\times10^{-10}$ (line 1),
$\epsilon_{0}=3.47\times10^{-10}$ a.u. (line(2),
$\epsilon_{0}=10.4\times10^{-10}$ a.u.(line 3). $10^{-9}$ a.u. on
the frequency axis corresponds to 1.2 $\Gamma$. }
%\label{f4}
\end{figure}


\newpage
\begin{thebibliography}{11}
\bibitem{c1}
M. O. Scully and M. S. Zubairy, {\em Quantum Optics}, (Cambridge
University Press, 1997).
\bibitem{c2}
C. Liu, Z. Dutton, C. H. Behroozi and L. V. Hau, Nature {\bf 409},
490 (2001).
\bibitem{c3}
D. F. Phillips, A. Fleischhauer, A. Mair, R. L. Walsworth and M.
D. Lukin, Phys. Rev. Lett. {\bf 86}, 783 (2001).
\bibitem{c4}
M. Fleischhauer and M. D. Lukin, Phys. Rev. Lett. {\bf 84}, 5094
(2000).
\bibitem{c5}
O. Kocharovskaya, Y. Rostovtsev and M. O. Scully, Phys. Rev. Lett.
{\bf 86}, 628 (2001).
\bibitem{c6}
S. E. Harris, Phys. Today {\bf 507}, 36 (1997).
\bibitem{d1}
A. B. Matsko, O. Kocharovskaya, Y. Rostovtsev, G. G. Welch, A. S.
Zibrov and A. O. Scully, Adv. At. Mol. Phys. {\bf 46}, 191 (2001).
\bibitem{d2}
Z. Dutton, N. S. Ginsberg, C. Slowe and L. V. Hau, Europhysics
News {\bf 35}, 33 (2004).
\bibitem{c7}
W. Demtr\"{o}der, {\em Laser Spectroscopy}, (Springer, Berlin,
1996).
\bibitem{c8}
G. S. Agarwal and Tarak Nath Dey, Phys. Rev. A{\bf 68}, 063816
(2003).
\bibitem{c9}
G. W\c{a}sik, W. Gawlik, J. Zachorowski and Z. Kowal, Phys. Rev. A
{\bf 64} 051802(R) (2001).

\end{thebibliography}
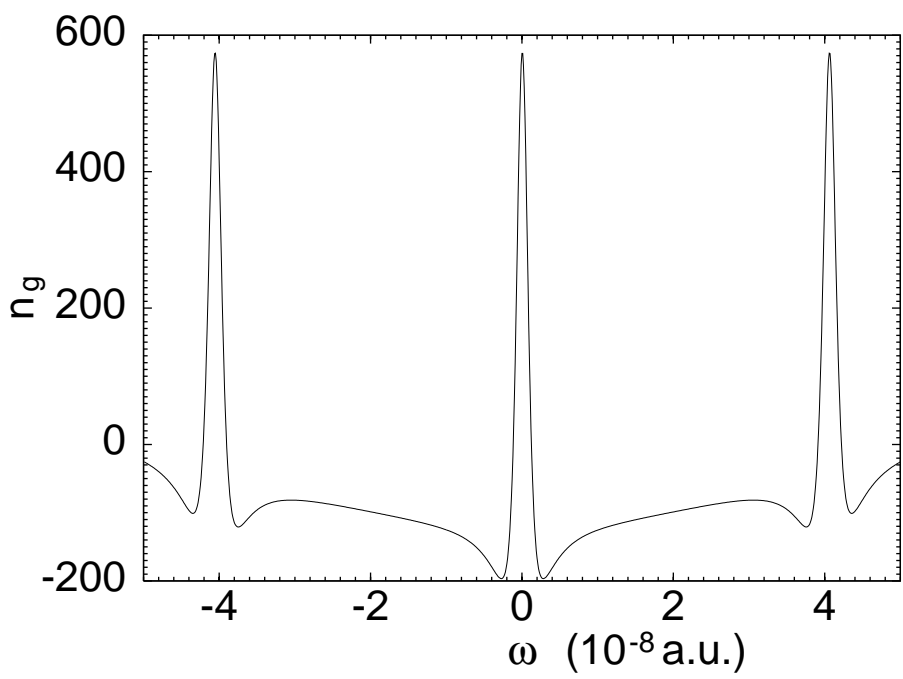
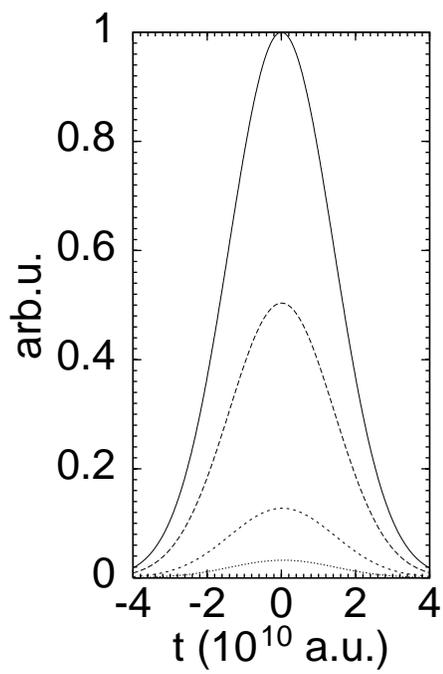
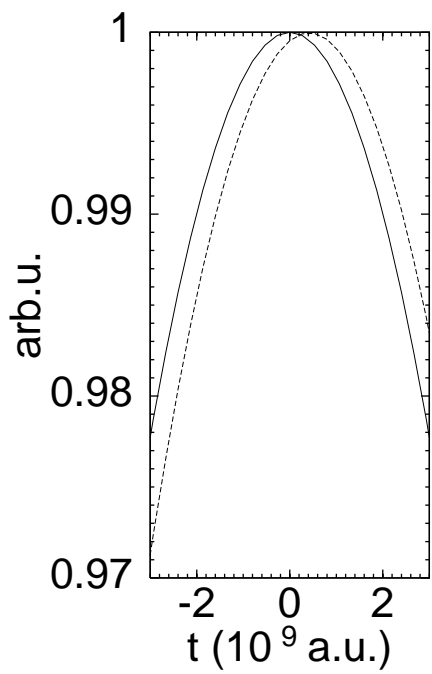
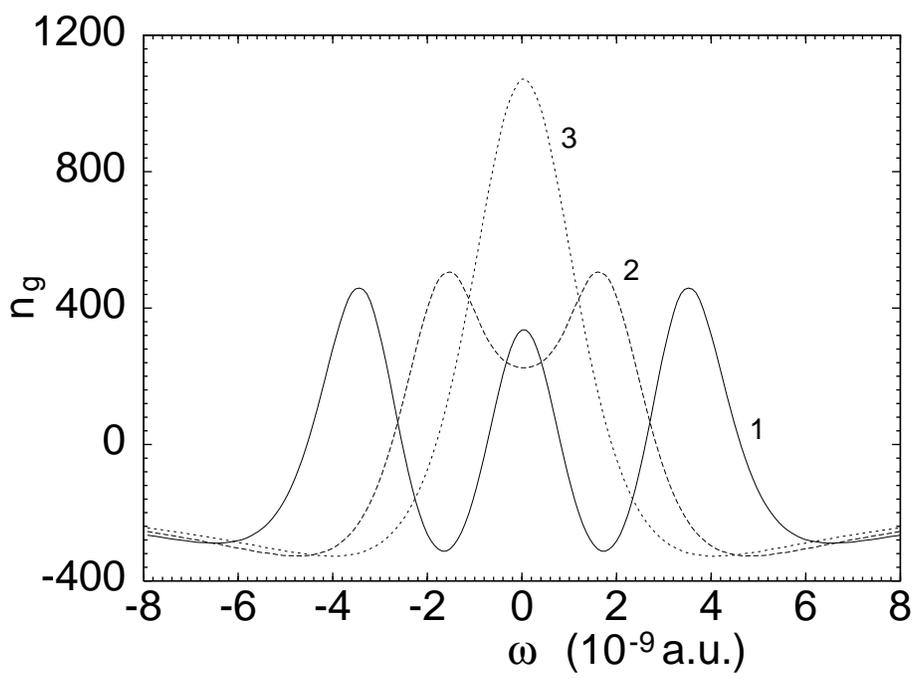
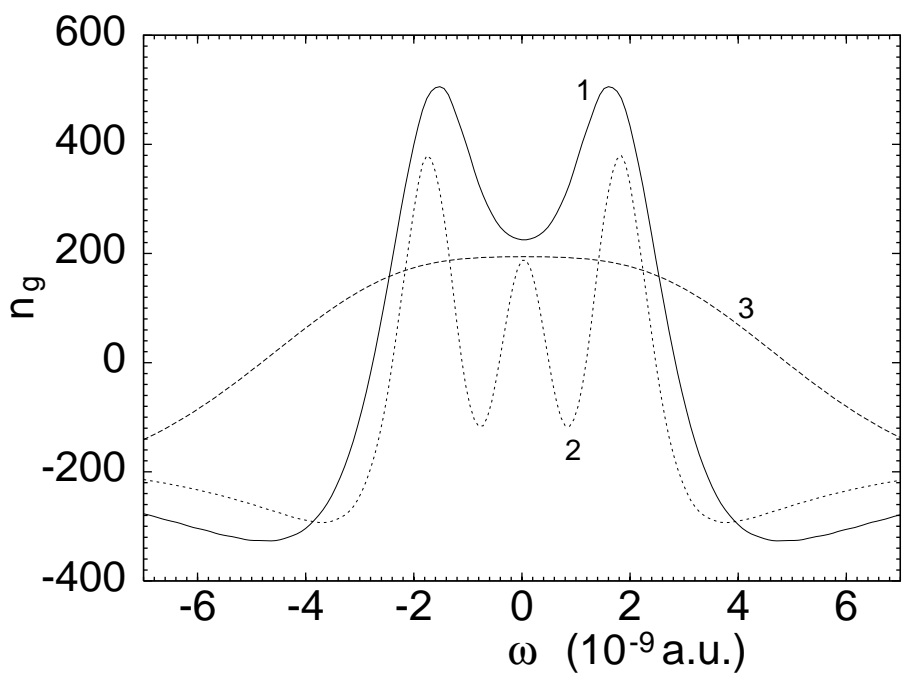
\end{document}